# The Possible Existence of Hs in Nature from a Geochemical Point of View

A. V. Ivanov

*Institute of the Earth's Crust, Siberian Division of the Russian Academy of Sciences, Irkutsk, Russia*
*e-mail: aivanov@crust.irk.ru*

**Abstract**

A hypothesis of the existence of a long-lived isotope $^{271}$Hs in natural molybdenites and osmirides is considered from a geochemical point of view. It is shown that the presence of Hs in these minerals can be explained only by making an additional *ad hoc* assumption on the existence of an isobaric pair of $^{271}$Bh-$^{271}$Hs. This assumption could be tested by mass-spectrometric measurements of U, Pb, Kr, Xe, and Zr isotopic shifts.

*Extraordinary claims require extraordinary evidence.*
*Carl Sagan*

## 1. INTRODUCTION

Calculated half lives of a hypothetical isotope $^{271}$Hs from $1.3\times10^{8}$–$1.8\times10^{11}$ yr depending on different adopted nuclei deformation parameters were reported recently in [1]. The aim of these calculations was to fit the theoretical data to the indications of ~4.4 MeV α activity experimentally found in some natural objects [2–4]. The possibility of Hs occurring in molybdenite and osmiride, in which this excess α activity was observed [2, 4], is considered from a geochemical point of view in the present paper. The problem of the existence of long-lived superheavy elements is discussed from the standpoint of theoretical and experimental physics, among other papers, in [5–9].

## 2. THE HISTORY OF THE QUESTION

In the early 1960s V. V. Cherdyshev and coauthors [10, 11] observed using the α-spectroscopic method an excess of 235U in magnetite and molybdenite, which necessitates the presence in nature of a transuranium radioactive element. Later works were aimed at the identification of this transuranium element with the use of the α-spectroscopic method. In papers [2–4] in natural samples of different age and genesis (bone fossils, magmatic minerals, iron meteorite, etc.) unidentified α activity in the energy range from 4.2–4.6 MeV was observed, as well as the presence of an energy spectrum identified with 239Pu was recognized. Besides, the $^{243}$Am energy spectra were detected in a number of samples. Papers [2, 3] made an assumption that 4.2–4.6 MeV α activity is due to $^{247}$Cm, which through two β$^-$ and two α decays turns into $^{239}$Pu, whereas 247 Cm by itself is a decay product of an even heavier transuranium element. The chemical properties of the transuranium element were similar to osmium [3]. Different fractions of osmiridium after chemical sample preparation were analyzed, as well as αactivity of energy 4.4 MeV was revealed in [4]. To account for the observed phenomenon, it was suggested that there possibly exists a long-lived superheavy element with the chemical properties of osmium, now known as hassium (Hs, element 108). The suggested chain of radioactive transformations between the hypothetical longlived isotope $^{271}$Hs and $^{247}$Cm was the following:

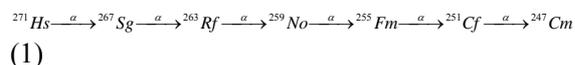

(1)

Through successive α and β$^-$ decays, $^{247}$Cm turns to a stable isotope $^{207}$Pb with an intermediate long-lived isotope $^{235}$U ($T_{1/2}$ = (703.8±0.5)×10$^6$ yr [12]):

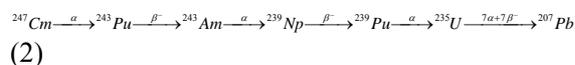

(2)

Based on the estimated abundance of the ~4.4 MeV α particle emitter, in [2] the $^{247}$Cm half-life was calculated to be (2.5±0.5) ×10$^8$ yr, which is an order of magnitude greater than the experimental value (1.56±0.05) ×10$^7$ yr [12]. The half-life 2.5±0.5) ×10$^8$ yr was later ascribed to $^{271}$Hs [1]. The results of [2–4, 10, 11] and the consequences emerging from them, even though important, were not verified at a later time. At the same time evidence for the observation of α activity in the ~4.4 MeV region in the thorite of Conway (Great Britain) granite [13] were not confirmed by the later investigations of the same samples [14, 15].

## 3. POSSIBLE PRESENCE OF Hs IN MOLYBDENITE AND OSMIRIDE

It is common knowledge that Hs is a homologue of Os [16]. This allows one to consider the possible presence of Hs in natural samples by analogy with Os abundance in them.

Let us consider molybdenite and osmiride, the minerals that were examined by the α-spectroscopic method in [2, 4]. Geological descriptions of the sampling points of the minerals studied are extremely scanty. For magmatic rocks, however, these minerals are typomorphic, which enables the known characteristics of these minerals to be extended to the samples under study.

Molybdenite (MoS$_2$) crystallizes in a subordinate amount in some kinds of acid (granite) magmas. It sometimes forms aggregates in granites, but it is more frequent in the zones of hydrothermal conditions. Molybdenite as a whole can be treated as a mineral, whose formation is due to the processes going on in the earth's crust. It is widely used for Re–Os dating; therefore, the distribution of Re and Os in it has been much studied [17–21 and the references therein]. The fact that this mineral from the outset does not contain osmium immediately attracts our attention. All osmium in it is represented by an isotope $^{187}$Os, which collects in the course of geological time through the β$^-$ decay of $^{187}$Re. On this basis it seems unlikely that molybdenite could capture any amount of hassium of importance, if this element occurs naturally. Thus, one should either call into question the results of papers [2–4, 10, 11] or infer that the isotope $^{271}$Hs is the decay product of an isotope of some other superheavy element. A hypothetical long-lived β$^-$ active isotope $^{271}$Bh (bohrium, element 107) could be such an isotope. Taking into account that Bh is a homologue of Re [22], while Re is usually contained in molybdenite in large quantities (tenth fractions of percent), one might expect that Bh, if any is in nature, will be picked up by the crystal lattice of this mineral during the formation of molybdenite.

To put it differently, the presence of Hs in molybdenites suggests the existence of sufficiently long-lived β$^-$ active $^{271}$Bh or of $^{271}$Sg (seaborgium, element 106), molybdenum's homologue, which through two β$^-$ decays goes into $^{271}$Hs. If this assumption is true, then the Hs presence in molybdenite finds a logical explanation from the viewpoint of the geochemical properties of this mineral.

Osmiride is a natural osmium–iridium alloy admixed with other platinoids, which is characteristic of ultrabasic rocks (peridotites). It is associated with the processes taking place in the Earth mantle [23–25 and the references therein]. Considering that Os is the basic element in this mineral, one should expect that Hs, if it exists in the Earth's mantle, must be present in this mineral in significant amount along with Os.

Assuming that the estimate of the $^{271}$Hs half-life $(2.5\pm0.5)\times10^8$ yr is true [2], then, essentially, all the $^{271}$Hs brought to the Earth during its formation, about 4.5 billion years ago, would have to decay. Superheavy elements are generated by supernova explosions and might be present in high-energy cosmic rays [26]. This suggests the existence of a continuous inflow to the Earth of superheavy elements, together with cosmic dust, which builds up as it traverses the solar system through the spiral arms of our galaxy. In an indirect way this fact may be indicated by $^{239}$Pu found in iron-manganese concretions on the floor of the Pacific Ocean and the Gulf of Finland [3]. Superheavy elements could penetrate the Earth's mantle through the subduction of oceanic sediments (immersion of an oceanic plate into the Earth's mantle in the descending branch of the mantle convection). This process of oceanic sediment entrance into the mantle was well documented through geological-geophysical and geochemical investigations [27–29 and the references therein].

Considering other ways Hs may end up in the Earth's mantle, let us assume that Hs is a product of $^{271}$Bh $\beta^-$ decay. According to current geochemical concepts, Re is concentrated principally in the Earth's metallic core [30, 31]. From this it may be suggested that the core is enriched with $^{271}$Bh and its decay product is $^{271}$Hs. Some authors are of the opinion that the core matter, enriched with the radiogenic isotope $^{187}$Os (and correspondingly with $^{271}$Hs, according to our assumption), may be brought to the Earth's surface by ascending convective fluxes (plumes) [32]. Such an interpretation was proposed, in particular, for some osmirides [33]. Other authors deny the possibility in principle for matter exchange between the core and the upper mantle [34, 35] or suppose only a limited exchange [36]. An enrichment of volcanic rocks and osmirides with the radiogenic $^{187}$Os may also be attributed purely to the processes taking place in the upper mantle [37] (see an overview of the problem [38]).

## 4. DISCUSSION

From the preceding section it is seen that the hypothesis of the presence of Hs in natural minerals molybdenite and osmiride cannot be accepted without a number of additional ad hoc assumptions. Moreover, the decay chain (equation (1)) suggested in [4] might not be found in nature at all, since by the new data, $^{263}$Rf decays by way of a spontaneous fission and not an α decay [12, 39]. Yet, accepting the hypothesis of the coexistence of long-lived isotopes Hs and Bh in molybdenite, one may assume the following chain of radioactive transformations:

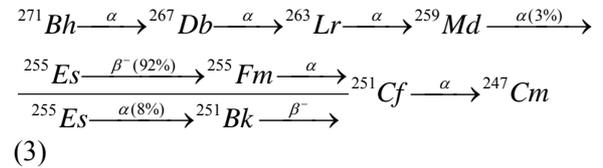

(3)

In this case the major portion of the decay products is made up from spontaneous fission fragments:

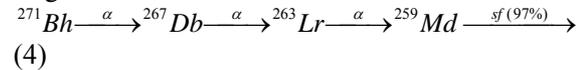

(4)

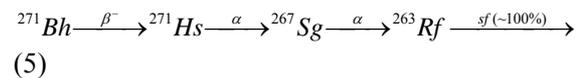

(5)

The isotopes $^{271}$Bh, $^{271}$Hs, $^{267}$Sg, $^{267}$Db, and $^{263}$Lr have not yet been produced experimentally, which is why the decay chains (3), (4), and (5) seem highly hypothetical. Thus, the idea of the presence of Hs long-lived isotopes in some natural objects [2–4] associated with the observed excess activity of $^{235}$U, $^{239}$Pu, and unidentified ~4.4-MeV ⟨ activity in various natural samples cannot be accepted without additional assumptions. In view of the fact that the data of the papers [2–4, 10, 11] were not verified experimentally, such a check is the barest necessity. In the natural samples enriched with the hypothetical isotope $^{271}$Hs it is good to expect an anomalous enrichment with the isotope $^{235}$U, while in the ancient samples (hundred millions - billions of years) one might expect, in addition to the above, an enrichment with $^{207}$Pb. Besides, isotope shifts must be observed for Kr, Xe, Zr, and other elements originated during spontaneous fission. Prior to the verification of these effects by direct mass-spectroscopic methods, attempts to interpret the results of [2–4, 10, 11] in the context of the existence of superheavy elements in nature are likely to be inadvisable.

## 5. CONCLUSIONS

A hypothesis, discussed in the literature [1–4, 10, 11], of the existence in natural samples of the longlived isotope Hs, based on α-spectrometric experimental data, has been considered. From the geochemical properties of Os (homologue of Hs) and Re (homologue of Bh) it has been concluded that these experimental data can be accounted for by making an assumption on the existence of the isobaric pair $^{271}$Bh–$^{271}$Hs. This assumption must be verified by measuring the isotope shifts of U, Pb, Kr, Xe, and Zr in natural objects (for example, molybdenites) by way of direct mass-spectroscopic measurements.

## ACKNOWLEDGMENTS

The work was supported by the Russian Foundation for Basic Research (grant no. 05-05-64281). The authoris grateful to A.A. Baldin for his support and V.V. Kobychev for helpful remarks.

## REFERENCES

1. A. Marinov et al., "New Outlook on the Possible Existence of Superheavy Elements in Nature," Phys. At. Nucl. **66**, 1137–1145 (2003).
2. V. V. Cherdyntsev and V. F. Mikhailov, "Primitive Transuranium Isotop in Nature," Geokhimiya, No. 1, 3–15 (1963).
3. V. V. Cherdyntsev et al., "Plutonium-239 in Nature," Geokhimiya, No. 4, 395–401 (1968).
4. H. Meier et al., "Über die Existenz einer Unbekannten Natürlichen α-Aktivität im 4.3–4.6 MeV Bereich," Z. Naturforsch. **25**, pp. 79–87 (1970).
5. Yu. Ts. Oganesyan, "Can Superheavy Elements Exist in Nature," Kratk. Soobshch. OIYaI, No. 6, 49–58 (1996).
6. Yu. Oganessian, "Synthesis and Decay Properties of Superheavy Atoms in Nuclear Reactions Induced by Stable and Radioactive Ion Beams," Eur. Phys. J. A **13**, 135–141 (2002).
7. A. Sobiczewski, "Present View of Stability of Heavy and Superheavy Nuclei," Usp. Fiz. Nauk **166**, 943–948 (1996) [Phys. Usp. **39**, 885–889 (1996)].
8. S. Hofmann, "Techniques for the Discovery of New Elements," Nucl. Instrum. Methods Phys. Res. B **126**, 310-315 (1997).
9. G. Royer, "Alpha Emission and Spontaneous Fussion through Quasi-Molecular Shapes," J. Phys. G: Nucl. Part. Phys. **26**, 1149–1170 (2000).
10. V. V. Cherdyntsev et al., "235U Excess in Magnetite with Excessive Content of Actinium," Geokhimiya, No. 4, 373–374 (1960).
11. V. V. Cherdyntsev et al., "Uranium Isotop in Natural Conditions. II. Isotopic Composition of Minerals," Geokhimiya, No. 10, 840–848 (1961).
12. T. V. Galashvili, *Handbook of Nuclides-2* (TsNIIATOMINFORM, Moscow, 2002) [in Russian].
13. R. D. Cherry, K. A. Richardson, J. A. S. Adams, "Unidentified Excess Alpha-Activity in the 4.4-MeV Region in Natural Thorium Samples," Nature **202**, 639–641 (1964).
14. K. A. Petrzhak, M. I. Yakunin, and G. M. Ter-Akopyan, "To the Problem of the α-Activity of Thorium," At. Energiya **32**, 179–181 (1972).
15. R. Gentry et al., "Reinvestigation of the α-Activity of Conway Granite," Nature **273**, 217–218 (1978).
16. Ch. E. Düllmann et al., "Chemical Investigation of Hassium (Element-108)," Nature **418**, 859–862 (2002).
17. A. V. Ivanov and S. V. Palesskii, "Analysis of osmium isotope ratios by ICP-MS on chemical etching of molybdenite as applied for Re-Os dating with preliminary neutron activation," Geochemistry International **41**, 1028-1032 (2003).

18. J. G. Raith and H. J. Stein, "Re–Os Dating and Sulfur Isotope Composition of Molybdenite from Tungsten Deposits in Western Namaqualand, South Africa: Implications for Ore Genesis and the Timing of Metamorphism," Mineral. Deposita **35**, 741–753 (2000).
19. D. Selby et al., "Re–Os and U–Pb Geochronology of the Clear Creek, Dublin Gulch, and Mactung Deposits, Tombstone Gold Belt, Yukon, Canada: Absolute Timing Relationships Between Plutonism and Mineralization," Can. J. Earth Sci. **40**, 1839–1852 (2003).
20. H. J. Stein et al., "The Remarkable Re–Os Chronometer in Molybdenite: How and Why it Works?" Terra Nova **13**, 479–486 (2001).
21. H. J. Stein et al., "Re–Os Ages for Archean Molybdenite and Pyrite, Kuittila-Kivisio, Finland and Proterozoic Molybdenite, Kibeliali, Lithuania: Testing the Chronometer in a Metamorphic and Metasomatic Setting," Mineral. Deposita **33**, 329–345 (1998).
22. R. Eichler et al., "Chemical Characterization of Bohrium (Element 107)," Nature **407**, 63–65 (2000).
23. A. Meibom, and R. Frei, "Evidence for an Ancient Osmium Isotopic Reservoir in Earth," Science **296**, 516–518 (2002).
24. A. Meibom, R. Frei, and N. H. Sleep, "Osmium Isotopic Compositions of Os-Rich Platinum Group Element Alloys from the Klamath and Siskiyou Mountains," J. Geophys. Res. B **109**, 02203 (2004).
25. A. Meibom et al., "Re–Os Isotopic Evidence for Long-Lived Heterogeneity and Equilibration Processes in the Earth's Upper Mantle," Nature **419**, 705–708 (2002).
26. V. P. Perelygin et al., "On Search and Identification of Tracks Due to Short-Lived SHE Nuclei in Extraterrestrial Crystals," Radiat. Meas. **36**, 271–279 (2003).
27. J. Morris, R. Valentine, and T. Harrison, "Be-10 Imaging of Sediment Accretion and Subduction Along the Northeast Japan and Costa Rica Convergent Margins," Geology **30**, 59–62 (2002).
28. J. P. Davidson, "Lesser Antilles Isotopic Evidence of the Role of Subducted Sediment in Island-Arc Magma Genesis," Nature **306**, 253–256 (1983).
29. W. M. White and B. Dupre, "Sediment Subduction and Magma Genesis in the Lesser Antilles ? Isotopic and Trace-Element Constraints, J. Geophys. Res. B **91**, 5927–5941 (1986).
30. R. J. Walker, J. W. Morgan, and M. F. Horan, "Os-187 Enrichment in Some Plumes—Evidence for Core-Mantle Interaction," Science **269**, 819–822 (1995).
31. R. J. Walker et al., "Applications of the Pt-190–Os-186 Isotope System to Geochemistry and Cosmochemistry," Geochim. Cosmochim. Acta **61**, 4799–4807 (1997).
32. A. D. Brandon et al., "Os-186–Os-187 Systematics of Hawaiian Picrites," Earth Planet. Sci. Lett. **174**, 25–42 (1999).
33. J. M. Bird et al., "Osmium and Lead Isotopes of Rare OsIrRu Minerals: Derivation from the Core-Mantle Boundary Region?" Earth Planet. Sci. Lett. **170**, 83–92 (1999).
34. D. L. Anderson, "Top-Down Tectonics," Science **293**, 216–218 (2001).
35. G. R. Foulger and J. H. Natland, "Is 'Hotspot' Volcanism a Consequence of Plate Tectonics," Science **300**, 921–922 (2003).
36. S. V. Balyshev and A. V. Ivanov, "Low-density anomalies in the mantle: ascending plumes and/or heated fossil lithospheric plates," *Doklady Earth Sciences* **380**, 858-862 (2001).
37. A. D. Smith, "Critical Evaluation of Re–Os and Pt–Os Isotopic Evidence on the Origin of Intraplate Volcanism," J. Geodynamics **36**, 469–484 (2003).
38. A. V. Ivanov and S. O. Balyshev, "Mass Flux across the Lower-Upper Mantle Boundary: Vigorous, Absent or Limited?" in *Plates, Plumes and Paradigms*, Ed. by G. R. Foulger et al. (Geological Society of America, Princeton, 2005), Spec. Paper 388, pp. 327–346.
39. J. V. Kratz et al., "An EC-Branch in the Decay of 27-s $^{263}$Db: Evidence for the Isotope $^{263}$Rf," Radiochim. Acta **91**, 59–62 (2003).